# A spatial analysis of COVID-19 reported cases in the Gauteng province, South Africa: Identifying wards to be targeted early in future infectious diseases outbreak

Mahdi Salehi [1,2], Samuel Manda [2], Mohammad Arashi [2,3], Andriette Bekker [2]

[1] Department of Mathematics and Statistics, University of Neyshabur, Iran; *salehi2sms@gmail.com*
[2] Department of Statistics, Faculty of Natural and Agricultural Science, University of Pretoria, Pretoria, South Africa; *samuel.manda@up.ac.za*; *andriette.bekker@up.ac.za*
[3] Department of Statistics, Faculty of Mathematical Science, Ferdowsi University of Mashhad, Mashhad, Iran; *arashi@um.ac.ir*

*Abstract*

The COVID-19 pandemic caused major disruptions and contributed to the loss of livelihoods and income. The pandemic also provided public health and health systems policy shifts towards better promotion and protection in responding to such disasters and emergencies. Due to differing effects of socio-economic infectious disease vulnerabilities and pre-pandemic levels of preparedness for health emergencies, health system strengthening requires targeted and ununiform implementation. We employ spatial statistical methods on the COVID-19 confirmed cases in identifying wards that could be targeted for strengthening health security in the Gauteng Province, South Africa. In this way, the identified high-risk wards would be more effective and prepared to respond to future pandemics and emergencies.

*Keywords:* COVID-19, Moran's index, LISA map, spatial regression models, ward level

**Introduction**

The restrictions imposed due to the COVID-19 pandemic caused devasting effects on the local and national economies, with severe losses in employment and income in Southern Africa (Strauss et la, 2021; Ngarava et al, 2022) . The widespread and severity of the economic impact of the pandemic varied by country depending on several factors including infectious disease vulnerability levels, pre-existing social and economic inequities, existing pandemic management plans and health systems (Manda et al, 2021; Rispel et al, 2021). The differential impact of the COVID-19 pandemic showed how critical a strong and robust health system is to contain and prevent new infections.

The observed variations in responses that resulted in the differential impact of the pandemic at the country level masked the same response variations within countries and were also shaped by lower-level diverse political, economic and health system infrastructure among an array of factors.  For example, South Africa, with a population of 60.1 million, has nine provinces whose health departments are semi-autonomous. However, there are great variations in resource availability, human resource capacity, and leadership that could have resulted in differential responses to the pandemic. The resource-constrained public health sector provides health care to 83% of the population. The remaining 17% of the population is covered by the private health insurance sector, which varies both by province with the urban provinces of Gauteng and the Western Cape covered most. Even within provinces, deep social economic impact of COVID-19 has had deep socio-economic vary by socio-economic and demographic group, with poorer households and Black Africans heavily impacted as was in the Gauteng province (Maree et al, 2021).

Following Maree et al (2021) who found a differential impact of COVID-19 at the municipal level in the Gauteng province, South Africa, we set out to identify lower administrative division (wards) that were hard hit by COVID-19 using spatial statistics methods on reported and confirm publicly available COVCID-19 cases. The Gauteng province contributes a third of South Africa's gross domestic product, making it the nation's biggest provincial economy. Thus, being the industrial and economic powerhouse of South Africa, disruptions due to the emergence and re-emergence of infectious diseases pose a substantial threat to the country's economy. Thus empirical knowledge and evidence of the most vulnerable areas will enable quicker and more targeted responses to mitigate the adverse impact of future epidemics on the economy.

Spatial statistics analyses have previously been used to support policy responses in the allocation of resources and COVID-19 control measures. For example, Liu et al (2021) performed a spatial clustering analysis of COVID-19 and its associated factors in mainland China at the regional level. By using spatial analysis, Mollalo et al (2021) emphasize future vaccine allocation and more medical professionals and treatment equipment should be a priority to those high-high clusters in the USA. Other similar spatial analyses of COVID-19 could be found in Jang et al (2021), Dutta et al (2021), Wang et al (2021), Golinelli et al (2021), Hernandez et al (2021), Salehi et al (2021), Arashi et al (2021), Alcântara et al (2020), Charlie and Gray (2020) and Zhao et al (2020), Mahmoudi et al (2021),Manda et al (2021), Sharma et al (2023), Kim et al (2023), Choiruddin et al. (2023) and Cribari-Neto (2023) Most of these used a first or second layer of administrative divisions, which could hinder effective responses at the lowest levels. By using a ward as a unit of analysis, we believe that responses to future pandemic will have improved preparedness due to the experience from COVID-19 as was the case with South East countries having had previous epidemics such as severe acute respiratory syndrome (SARS) and influenza A virus (N1H1) (Nit et al, 2021; Nguyen, et al, 2021).

**Study Area**

South Africa contains a high-risk population combined with low-income country characteristics (Arashi et al, 2021, Schlüter et al. 2021). Though Gauteng is the smallest of South Africa's nine provinces by size, it is the most populous (15 million people) and wealthiest province, and center of the country's economy, powerhouse contributing a third of its gross domestic product (GDP). Having the smallest land area and the largest population in South Africa, the province of Gauteng has the highest population density, but most people are clustered in major metropolitan areas of Johannesburg, Tshwane and Ekurhuleni and on the edges of the urban areas. These disparities in population densities and crowdedness could impact COVID-19 levels between locations and wards in the province. Administratively, the province is divided into three metropolitan municipalities and two district municipalities each district municipality is in turn divided into six local municipalities. Thus, the Gauteng province has nine (9) municipalities (level 3 administrative division). The metropolitan and local municipalities are further divided into about 529 wards. Figure 1 and Table 1 show a map of the Gauteng province and the number of wards that falls within each municipality, respectively.

We investigated the spatial distribution of the COVID-19 ecological predictors and the number of COVID-19 confirmed cases. Moreover, we obtained the global and local Moran's indexes of the latter variable based on various types of weights in order to find the possible clusters among 529 wards of Gauteng given by Table 1. We found that the Municipalities Johannesburg and Ekurhuleni have had the most contribution in the number of hotspot wards as they had the most confirmed cases in comparison with the remaining municipalities. This retrospective study emphasizes the importance of data modeling at micro levels in order to timeously predict any major outbreaks. To this end, in addition to the ordinary least square (OLS) model, we also employed the spatial lag model (SLM), the spatial error model (SEM) as well as the geographical weighted regression (GWR) to predict the COVID-19 prevalence in Gauteng. The results show that there is no longer a significant auto-correlation for the residuals of the spatial models, as a result, employing such spatial regression models is necessary here.

**Results**

**The spatial distributions of the variables of interest**

The main variable of interest in this study is the *number of COVID-19 confirmed cases* reported for each ward in Gauteng. Table 2 reports the relative frequency of the COVID-19 confirmed cases for each main sub-region of Gauteng given by Figure 1 and Table 1. Table 2 shows that Municipalities 2 and 9 have had the most confirmed cases compared to the other municipalities. Figure 2 shows more a high-spatial-resolution of the distribution of COVID-19 confirmed cases at ward level in the province. The central, as well as south-eastern wards, recorded a high number of cases compared to northern and eastern areas.

Figure 3 shows the spatial distribution of the average age of the positive cases reported for each ward in Gauteng (the right pane) and the distribution of this variable and the number of confirmed cases together (the right pane). As it is observed, the average age ranges from 16 to 68. More precisely, the first quartile, the median, and the third quartile of this variable are approximately equal to 38, 40 and 42 respectively. This means that the age of 50% of confirmed cases has fallen within the interval (38, 42) and the rest have been out of it.

Figure 4 shows the spatial distribution of the percentage of COVID-19 male patients (the left pane) and the joint spatial distribution of percentage of COVID-19 male patients and the number of confirmed cases (the right pane). Evidence shows that 50 percent of the wards have recorded at least 56.8% female positive cases.

The spatial distribution of the average location score associated with the confirmed cases in each ward of Gauteng as well as its joint distribution with the number of confirmed cases are reported in the left and the right panes of Figure 5, respectively. It is seen that the location score of the confirmed cases changes from 61.38 to 86.01 in average.

**The spatial association of the COVID-19 confirmed cases**

We have calculated the global Moran's I of the confirmed cases variable based on all weight matrices defined by Appendix (A). More specifically, we have used the 0-1 adjacency weight, $k$-nearest neighbor weight with $k = 5$ and 10, the distance-based weight with $d_0 = 15$ KM (the first decile of all distances) as well as the inverse distance weight with power coefficient $\alpha = 1$. The confirmed c*ases* variable does not follow from the normal distribution (the *Shapiro-Wilk p-value is less than 2.2e-16*); as a result, we have done the global Moran's *I* significance test under randomization. The estimated value, the variance, and the corresponding significance test of the global Moran's index of the confirmed c*ases* variable calculated based on different weight matrices are reported by Table 3.

Table 2 states that there is no significant spatial auto-correlation using the adjacency and 5-nearest neighbour weight matrices. On the other side, the rest of the weights show the existence of a **positive** spatial association for the confirmed cases variable. Therefore, there will be a sort of clustering for this variable over the municipalities of Gauteng province. Now, we intend to go a step further and find the possible clusters. To this end, we use the local Moran's index given by (2). Figure 6 displays the LISA cluster map of the municipalities of Gauteng for the confirmed cases variable. Indeed, this figure highlights the districts having passed the significance test of the local Moran for the mentioned variable. A significant positive local Moran's value for a specific district implies that it has similar values as its near districts, thus those sites constitute a spatial cluster. That spatial cluster can be a high–high cluster (high values surrounded by high-value neighbourhood) or a low–low cluster (low values in a low-value neighbourhood). On the other hand, a significant negative local Moran's value indicates that the site under study is a spatial outlier. Spatial outliers include high–low (locations with high values surrounded by regions with low values) and low–high (a low value in a high-value neighbourhood) outliers. As it is seen from all panes of Figure 6, there is a sort of high-high clustering located exactly at the central areas of Gauteng. However, four low-high outliers, as well as a high-low one, are detected based on the distance-based adjacency weight. In order to find the exact location of the clusters and those few outliers, refer to Table 4. It is observed from this table that the main sub-regions *Johannesburg* and *Ekurhuleni* (see Figure 1 and Table 1 for more information) have had the most contribution in the number of hotspot wards as they had the most confirmed cases in comparison with the remaining sub-regions (see Table 2). However, municipalities Mogale and Midvaal have also recorded one hotspot ward based on the inverse distance weight.

**The regression analysis**

In this section we intend to predict the COVID-19 prevalence (the number of COVID-19 confirmed cases reported per 1000 residents) in Gauteng's wards based on some regression models. Figures 7-8 exhibits the spatial and non-spatial distributions of this variable. Based on the histograms given by Figure 8, it is seen that the COVID-19 prevalence is not symmetrically distributed while its logarithm almost is. Thus, the latter dependent variable was used in the regression models. The geographically weighted regression (GWR) and the ordinary least square (OLS) as local and global models were utilized to determine the best fit for the logarithm of the COVID-19 prevalence in Gauteng's wards. Moreover, spatial lag regression models were also employed here. There are two types of spatial lag regression models; the spatial lag model (SLM), which models dependency in the response variable, and the spatial error model which models dependency in the residuals instead. For more details on the models used, refer to Appendix (B). The results of the models fitted are given by Tables 5-8. As it is observed from the OLS's results, the significant variables are as follows

'Population_density',
'No_flush_toilet_facilities',
'Hunger_risk',
'Failed_to_find_healathcare_when_needed'.

It is also concluded from the first row of Tables 7 and 8 that a spatial lag on the dependent variable and the residuals is needed. The criteria used for comparing different regression models were the $R^2$, the Akaike information's criterion (AIC) and the residual sum of squares (RSE). The resultant indexes are presented in Table 9. It is seen that the spatial regression models have better performance than the OLS. However, the GWR has outperformed the other counterparts based on all three criteria mentioned. Figure 9 displays the spatial distribution of the local $R^2$ obtained for each ward datum. The intuition shows that it varies from 0.17 to 0.22. Although the local values of $R^2$ obtained in the GWR are not that much acceptable, but still larger than that of the global model. As it is observed from Table 9, there is a significant spatial association for the residuals of the OLS model. As a result, employing spatial regression models for reducing such auto-correlation have been necessary. Table 9 confirms this fact: there is no longer a significant auto-correlation for the residuals of the spatial models. Moreover, the Moran's I statistic of the SEM is the lowest value among the other models. The same fact can result from the LISA plots of the residuals of the models displayed in Figure 10. As it is seen from this figure, SEM has recorded the minimum number of spatial clusters compared to the other counterparts.

## Conclusions and Discussions

Decision-making is the principle objective of COVID-19 studies. Therefore, these results from the retrospective study highlight the importance of investigating the impact of spreading and mitigation of disease on a prefecture level to make suggestions to local/federal government. This paper provided a framework to identify the social-economic and public factors that are the most influential contributors to risk at the individual sub-systems. We set out to identify high lowest administrative division in the Gauteng province which has the highest population density and the nation's biggest provincial economy in South Africa. We used COVID-19 confirmed cases and employed spatial statistics methods. Several infectious disease outbreaks and response variables were used in the analyses. More precisely, one of the main objectives of this study was to obtain an overall measure of spatial association for the confirmed cases variable. In this regard, the global Moran's I revealed that there exist a positive spatial association for the confirmed cases variable. Then, by using the local Moran index and its corresponding LISA map we realized that the municipalities Johannesburg and Ekurhuleni had the highest number of (High-High) hotspot wards as they also had the most confirmed cases compared with the other municipalities. In order to predict the prevalence of the COVID-19 of Gauteng's wards, we used 11 relevant explanatory variables measured for each wards. To this end, we utilized the OLS as a global regression model, the GWR as a local regression model and two types of spatial lag regression models including SEM and SLM. Results obtained based on various model selection criteria showed that the GWR outperforms the other competitors. Moreover, intuition suggested that the factors 'Population_density', 'No_flush_toilet_facilities', 'Hunger_risk', 'Failed_to_find_healthcare_when_needed' had the highest influence on the prediction of the prevalence of the COVID-19 in Gauteng.

## Appendices

### (A) Spatial auto-correlation

A spatial autocorrelation measures how distance influences a particular variable. In other words, it quantifies the degree of which objects are similar to nearby objects. Variables are said to have a positive spatial autocorrelation when similar values tend to be nearer together than dissimilar values, otherwise, variables are said to have a negative spatial autocorrelation when dissimilar values tend to be nearer together than similar values. An important step in spatial statistics and modelling is to get a measure of the spatial influence between geographic objects. This can be expressed as a function of adjacency or (inverse) distance, and is often expressed as a spatial weight matrix. See below different definitions of such weight matrices.

0-1 adjacency weight: $W = [w_{ij}]$, where $w_{ij} = 1$ if points $i$ and $j$ are

| k-nearest neighbour weight: | $W = [w_{ij}]$, where $w_{ij} = 1$ if the point $j$ is one of the $k$-nearest neighbour of the point $i$ and $w_{ij} = 0$ otherwise |
|---|---|
| Inverse distance weight: | $W = [w_{ij}]$, where $w_{ij} = \frac{1}{d_{ij}^{\alpha}}$ with power coefficient $\alpha$ and $d_{ij}$ stands for the distance between the points $i$ and $j$. |
| Distance-based adjacency weight: | $W = [w_{ij}]$, where $w_{ij} = 1$ if $d_{ij} < d_0$ and $w_{ij} = 0$, otherwise. |

adjacent and $w_{ij} = 0$ otherwise

The spatial association can be evaluated globally and locally. Global statistics summarize spatial autocorrelation, while local measures examine the individual locations, enabling hotspots to be identified based on comparisons to the neighboring samples. Moran's Index, originally defined by Moran (1950), has both of local and global representations. It has been defined as a measure of choice for scientists, specifically in environmental sciences, ecology and public health. Suppose that $y_1, \ldots, y_n$ are observations of the variable of interest and $W = [w_{ij}]$ is one of the above-mentioned weight matrices, then the global Moran's I is defined as follows

$$I = \frac{n}{S_o} \cdot \frac{\sum_{i=1}^{n}\sum_{j=1}^{n} w_{ij}(y_i - \bar{y})(y_j - \bar{y})}{\sum_{i=1}^{n}(y_i - \bar{y})^2}, \quad i \neq j \tag{1}$$

where $\bar{y} = n^{-1} \sum_{i=1}^{n} y_i$ and $S_o = \sum_{i=1}^{n}\sum_{j=1}^{n} w_{ij}$. The global Moran's I given by (1) varies from -1 to +1 and $I = 0$ shows no spatial association between the sub-regions for the underlying feature, i.e. the variable is distributed randomly over the region of study. But, the values of global Moran's $I$ near +1 indicate a significant positive association; as a result, there is a sort of clustering. On the other hand, the values close to -1 indicate strong negative spatial association which means that potential outliers exist. However, in order to ensure that there is a significant spatial auto-correlation, we test a hypothesis which is done under randomization or normality of the variable of the interest, say $y$. The null hypothesis states that $I = 0$, thus there is no spatial correlation, while the alternative hypothesis indicates the existence of a significant spatial association.

Note that measures of global spatial autocorrelation such as the one given by (1) do not reflect the local spatial correlations within geographic units. To tackle this problem, we employ the local Moran's $I_i$. Under the assumptions of this section, it is defined as

$$I_i = n \frac{(y_i - \bar{y}) \sum_{j=1}^{n} w_{ij}(y_j - \bar{y})}{\sum_{i=1}^{n}(y_i - \bar{y})^2}, \quad i = 1, \ldots, n. \tag{2}$$

It is clear that $\sum_{j=1}^{n} I_i = S_0 I$, where $S_0$ is the summation of the weights already defined in (1). A high positive local Moran's $I_i$ value implies that the location i has similarly high or low values as its neighbours, thus the locations are spatial clusters. Spatial clusters include high–high clusters (high values in a high value neighbourhood) and low–low clusters (low values in a low value neighbourhood). A high negative local Moran's I value means that the location under study is a spatial outlier. Spatial outliers are those values that are obviously different from the values of their surrounding locations. Spatial outliers include high–low (a high value in a low value neighbourhood) and low–high (a low value in a high value neighbourhood) outliers. It is beneficial to sketch a map which labels such relationships for each sub-region. Such a plot is so-called Anselin's Local Indicators of Spatial Association (LISA) cluster map.

**(B) Spatial regression models**

GWR is a regression model intended to indicate where locally weighted regression coefficients deviates from their global values obtained in the OLS model. Its basis is the concern that the fitted coefficient values of a global model e.g. the OLS, fitted to all the data, may not represent detailed local variations in the data (see Bivand et al, 2013). Thus, it can be considered as an extension of the ordinary OLS model since it involves the spatial locations of the data into the model as well. The GWR is expressed as follows (Fotheringham et al, 2002)

$$y_i = \beta_0(u_i, v_i) + \sum_{k=1}^{p} \beta_k(u_i, v_i) x_{ik} + \epsilon_i,$$

where $i = 1, \ldots, n$, denotes the $i$th location (observation), $\beta_0(u_i, v_i)$ is the constant at point $i$ and $\beta_k(u_i, v_i)$ is the $k$th local coefficient associated with the $k$th independent variable at location $i$ ($x_{ik}$).

The SLM and SEM are also extensions of the conventional OLS model. The former is formulated as follows

$$\boldsymbol{y} = \rho W \boldsymbol{y} + X\boldsymbol{\beta} + \boldsymbol{\epsilon},$$

where $\boldsymbol{y} = (y_1, \ldots, y_n)^T$ is the observations of the dependent variable, $W$ is the 0-1 adjacency weight matrix defined earlier, $\boldsymbol{\beta}$ is the column vector of the regression coefficients and $X$ is the design matrix. The SEM is defined as

$$\boldsymbol{y} = X\boldsymbol{\beta} + \boldsymbol{u},$$

where $\boldsymbol{u} = \lambda W \boldsymbol{u} + \boldsymbol{\epsilon}$.

**Data selection and handling**

The data of COVID-19 positive cases, their age, gender and location scale in Gauteng province are gathered since March 2020 to July 2020.

**Competing interests**

The authors declare no compering interest.

Ngarava S, Mushunje A, Chaminuka P, Zhou L. (2022). Impact of the COVID-19 pandemic on the South African tobacco and alcohol industries: Experiences from British American Tobacco and Distell Group Limited. Phys Chem Earth;127:103186. doi: 10.1016/j.pce.2022.103186. Epub 2022 Jun 16. PMID: 35757561; PMCID: PMC9212888.

Manda SOM, Darikwa T, Nkwenika T, Bergquist R. (2021). A Spatial Analysis of COVID-19 in African Countries: Evaluating the Effects of Socio-Economic Vulnerabilities and Neighbouring. *Int J Environ Res Public Health,* 14;18(20):10783. doi: 10.3390/ijerph182010783. PMID: 34682528; PMCID: PMC8535688.

Nit, B.; Samy, A.L.; Tan, S.L.; Vory, S.; Lim, Y.; Nugraha, R.R.; Lin, X.; Ahmadi, A.; Lucero-Prisno, D.E., 3rd. (2021). Understanding the Slow COVID-19 Trajectory of Cambodia. *Public Health Pract,* 2, 100073.

Nguyen Thi Yen, C.; Hermoso, C.; Laguilles, E.M.; De Castro, L.E.; Camposano, S.M.; Jalmasco, N.; Cua, K.A.; Isa, M.A.; Akpan, E.F.; Ly, T.P.; et al. (2021). Vietnam's success story against COVID-19. *Public Health Pract*., 2, 100132.

Maree, G., Culwick Fatti, C., Götz, G., Hamann, C. ,Parker, A. (2021). Effects of the COVID-19 pandemic on the gauteng city-region: findings from the GCRO's quality of life survey 6 (2020/21). Gauteng City-Region Observatory (GCRO). DOI: https://doi.org/10.36634/2021.db.2
11

**Tables and figures**

*Table 1. The number of wards falls within each Gauteng's sub-regions (municipalities).*

| Sub-region | Description | Number of wards |
|---|---|---|
| 1 | City of Tshwane | 107 |
| 2 | City of Johannesburg | 135 |
| 3 | Rand West City | 35 |
| 4 | Merafong City | 28 |
| 5 | Mogale City | 39 |
| 6 | Lesedi | 13 |
| 7 | Midvaal | 15 |
| 8 | Emfuleni | 45 |
| 9 | City of Ekurhuleni | 112 |

*Table 2. The distribution of the number of COVID-19 confirmed cases by municipality number in Gauteng.*

| **Sub-region** | **1** | **2** | **3** | **4** | **5** | **6** | **7** | **8** | **9** |
|---|---|---|---|---|---|---|---|---|---|
| **Frequency ratio** | 0.189 | 0.383 | 0.031 | 0.034 | 0.050 | 0.010 | 0.028 | 0.051 | 0.225 |

*Table 3: The Moran's I significance test for the number of COVID-19 confirmed cases of Gauteng's wards obtained based on various weight matrices.*

| Weight type | Estimate | Expectation | Variance | P-value |
|---|---|---|---|---|
| 0-1 Adjacency | 0.00906 | -0.00189 | 0.00044 | 0.29986 |
| 5-nearest neighbour | 0.02881 | -0.00189 | 0.00042 | 0.06699 |
| 10-nearest neighbour | 0.03633 | -0.00189 | 0.00021 | 0.00444 |
| Distance-based adjacency | 0.03065 | -0.00191 | 0.00005 | 0.00000 |
| Inverse distance | 0.01490 | -0.00189 | 0.00002 | 0.00014 |

*Table 4: Frequency table of hotspot wards discovered by the LISA plot (Figure 6) for each main sub-region displayed in Figure 1 and Table 1.*

| Weight matrix | | low-low | low-high | high-low | high-high |
|---|---|---|---|---|---|
| 0-1 adjacency | Municipality | - | - | - | **2  7  9** |
| | Frequency | 0 | 0 | 0 | 9  1  3 |
| 5-nearest neighbour | Municipality | - | - | - | 2 |
| | Frequency | 0 | 0 | 0 | 8 |
| 10-nearest neighbour | Municipality | - | - | - | **2  9** |
| | Frequency | 0 | 0 | 0 | 11  2 |
| Distance-based adjacency | Municipality | - | **1  5  7** | 1 | **2  9** |
| | Frequency | 0 | 2  1  1 | 1 | 22  6 |
| Inverse distance | Municipality | - | - | - | **2  5  7  9** |
| | Frequency | 0 | 0 | 0 | 20  1  1  9 |

*Table 5: Parameter summary of the global regression model (OLS) fitted on the prevalence of the COVID-19 of Gauteng's wards with some relevant explanatory variables.*

| | Estimate | Std. Error | P-value |
|---|---|---|---|
| (Intercept) | 2.1031 | 0.9504 | 0.0275* |
| Population_density | -3.238e-05 | 1.600e-05 | 0.0437* |
| Crowded_dwellings | -0.0180 | 0.0146 | 0.2190 |
| No_flush_toilet_facilities | -0.0212 | 0.0103 | 0.0393* |
| No_piped_water | 0.0107 | 0.0137 | 0.4369 |
| Zero_electronic_communication | 0.0425 | 0.0534 | 0.4272 |
| Poor_health_status | -0.0248 | 0.0221 | 0.2612 |
| No_medical_insurance | 0.0155 | 0.0096 | 0.1047 |
| Hunger_risk | -0.0275 | 0.0106 | 0.0100* |
| Pre_exisitng_health_conditions | 0.0021 | 0.0108 | 0.8435 |
| Difficult_to_save_money | 0.0068 | 0.0126 | 0.5916 |
| Failed_to_find_healthcare_when_needed | -0.0659 | 0.0320 | 0.0398* |

Table 6: Parameter summary of the local regression model (GWR) fitted on the prevalence of the COVID-19 of Gauteng's wards with some relevant explanatory variables

|  | Minimum | Q1 | Median | Q3 | Maximum |
|---|---|---|---|---|---|
| (Intercept) | 1.7628 | 1.7966 | 1.8681 | 1.8979 | 1.9204 |
| Population_density | -0.2451 | -0.2190 | -0.1920 | -0.1521 | -0.1302 |
| Crowded_dwellings | -0.2868 | -0.2640 | -0.2147 | -0.1619 | -0.1203 |
| No_flush_toilet_facilities | -0.5153 | -0.4793 | -0.4576 | -0.4134 | -0.3712 |
| No_piped_water | 0.1249 | 0.1977 | 0.2606 | 0.3024 | 0.3607 |
| Zero_electronic_communication | -0.1102 | -0.0458 | 0.0130 | 0.1130 | 0.1853 |
| Poor_health_status | -0.1700 | -0.0771 | -0.0498 | -0.0220 | 0.0293 |
| No_medical_insurance | 0.2616 | 0.3147 | 0.3751 | 0.4647 | 0.5249 |
| Hunger_risk | -0.8278 | -0.7785 | -0.7326 | -0.5699 | -0.4335 |
| Pre_exisitng_health_conditions | -0.0484 | -0.0239 | 0.0432 | 0.0853 | 0.1187 |
| Difficult_to_save_money | 0.0737 | 0.1087 | 0.1275 | 0.1478 | 0.1838 |
| Failed_to_find_healathcare_when_needed | -0.3730 | -0.3050 | -0.1859 | -0.1100 | -0.0350 |

Table 7: Parameter summary of the spatial lag model (SLM) fitted on the prevalence of the COVID-19 of Gauteng's wards with some relevant explanatory variables

|  | Estimate | Std. Error | P-value |
|---|---|---|---|
| $\rho$ | 0.0378 | 0.0121 | 0.0032 (LR test) |
| (Intercept) | 1.5688 | 0.9400 | 0.0951 |
| Population_density | -2.6342e-05 | 1.5620e-05 | 0.0917 |
| Crowded_dwellings | -0.0178 | 0.0142 | 0.2083 |
| No_flush_toilet_facilities | -0.0192 | 0.0100 | 0.0544 |
| No_piped_water | 0.0106 | 0.0133 | 0.4243 |
| Zero_electronic_communication | 0.0323 | 0.0518 | 0.5333 |
| Poor_health_status | -0.0208 | 0.0214 | 0.3311 |
| No_medical_insurance | 0.0155 | 0.0093 | 0.0952 |
| Hunger_risk | -0.0230 | 0.0104 | 0.0262 |
| Pre_exisitng_health_conditions | 0.0043 | 0.0104 | 0.6822 |
| Difficult_to_save_money | 0.0048 | 0.0122 | 0.6942 |
| Failed_to_find_healathcare_when_needed | -0.0629 | 0.0310 | 0.0422 |

*Table 8: Parameter summary of the spatial error model (SEM) fitted on the prevalence of the COVID-19 of Gauteng's wards with some relevant explanatory variables*

|  | Estimate | Std. Error | P-value |
|---|---|---|---|
| $\lambda$ | 0.0329 | 0.0143 | 0.0416 (LR test) |
| (Intercept) | 2.1679 | 0.9727 | 0.0258 |
| Population_density | -3.1986e-05 | 1.5878e-05 | 0.0440 |
| Crowded_dwellings | -0.0167 | 0.0146 | 0.2523 |
| No_flush_toilet_facilities | -0.0195 | 0.0102 | 0.0562 |
| No_piped_water | 0.0117 | 0.0135 | 0.3835 |
| Zero_electronic_communication | 0.0304 | 0.0522 | 0.5607 |
| Poor_health_status | -0.0219 | 0.0215 | 0.3079 |
| No_medical_insurance | 0.0146 | 0.0095 | 0.1225 |
| Hunger_risk | -0.0259 | 0.0106 | 0.0147 |
| Pre_exisitng_health_conditions | 0.0045 | 0.0107 | 0.6755 |
| Difficult_to_save_money | 0.0035 | 0.0126 | 0.7797 |
| Failed_to_find_healathcare_when_needed | -0.0563 | 0.0313 | 0.0715 |

*Table 9: Numerical indexes of the various regression models fitted on the prevalence of the COVID-19 of Gauteng's wards.*

|  | $R^2$ | AIC | RSE | Moran's I significance test | | | |
|---|---|---|---|---|---|---|---|
|  |  |  |  | Estimate | Expectation | Variance | P-value |
| OLS | 0.12 (adj) | 1680 | 1467 | 0.0594 | -0.0025 | 0.0009 | 0.0438 |
| GWR | **0.20** (Quasi) | **1649** | **1379** | 0.0507 | -0.0025 | 0.0009 | 0.0834 |
| SLM | - | 1673 | 1424 | -0.0105 | -0.0025 | 0.0009 | 0.7934 |
| SEM | - | 1677 | 1443 | **0.0031** | -0.0025 | 0.0009 | 0.8531 |

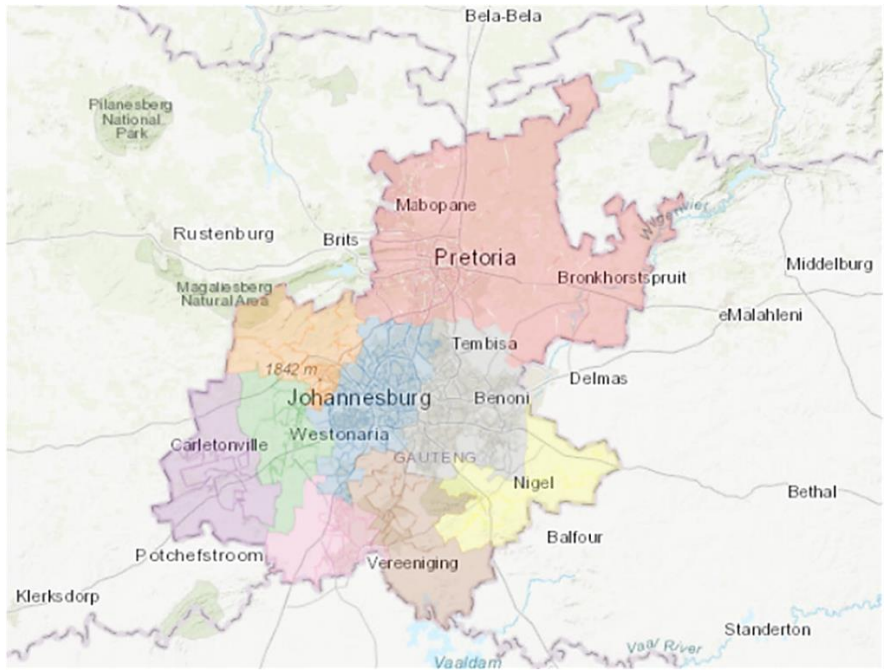

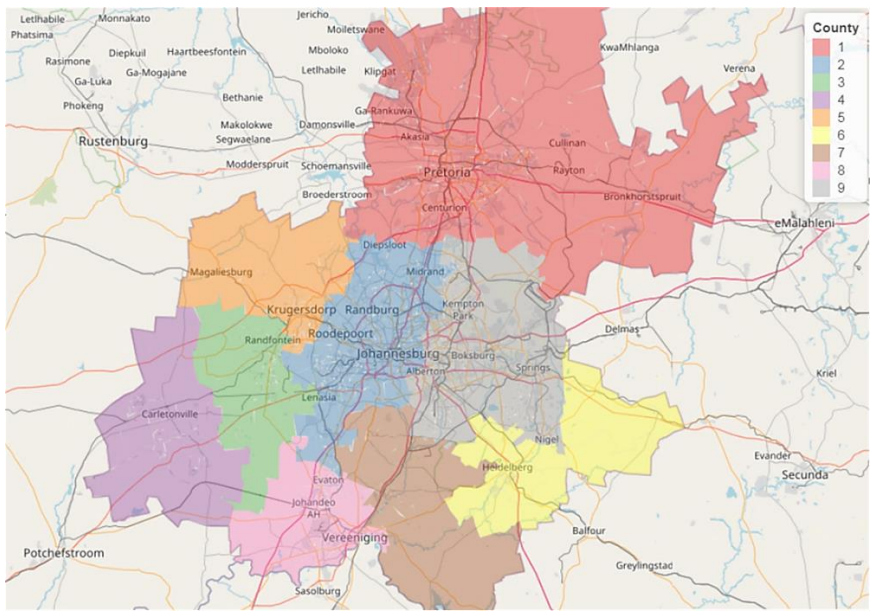

*Figure 1: Map displaying the nine municipalities in Gauteng province, South Africa as well as major connecting roads.*

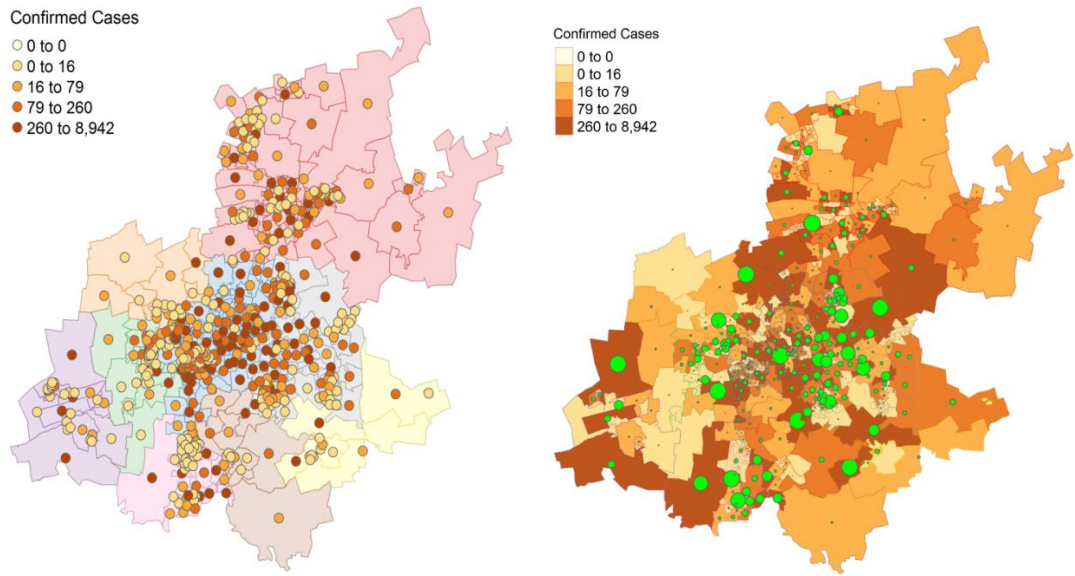

*Figure 2: The spatial distribution of COVID-19 confirmed cases in Gauteng province and its municipalities.*

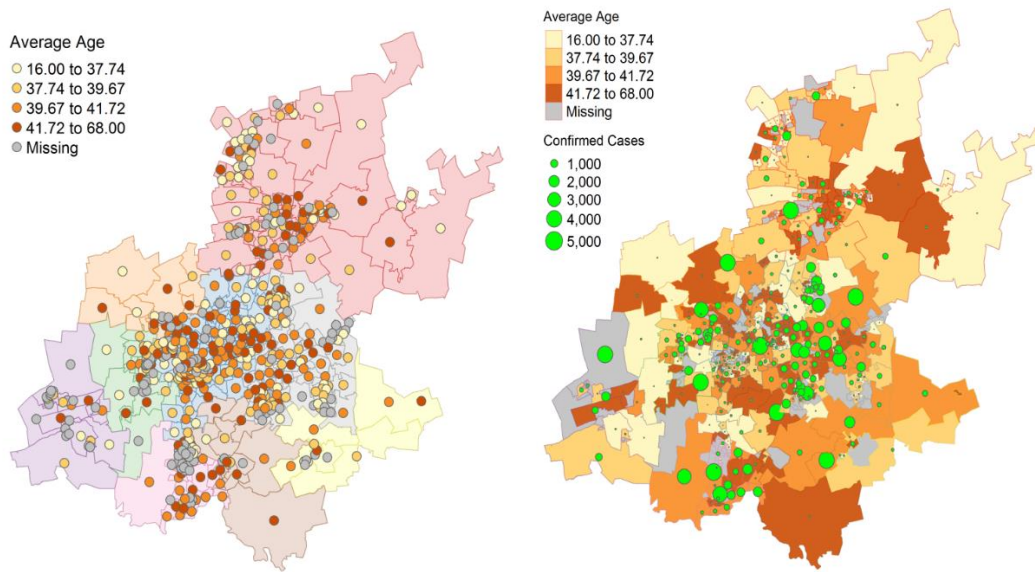

*Figure 3: The spatial distribution of the mean age of the COVID-19 confirmed cases, Gauteng.*

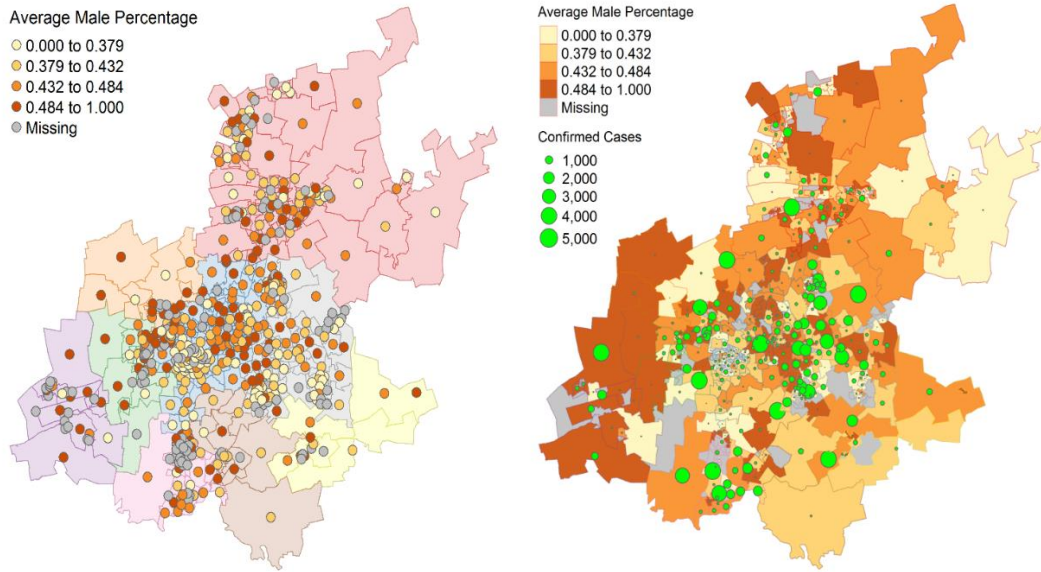

*Figure 4: The spatial distribution of percentage of the confirmed COVID-19 male cases, Gauteng.*

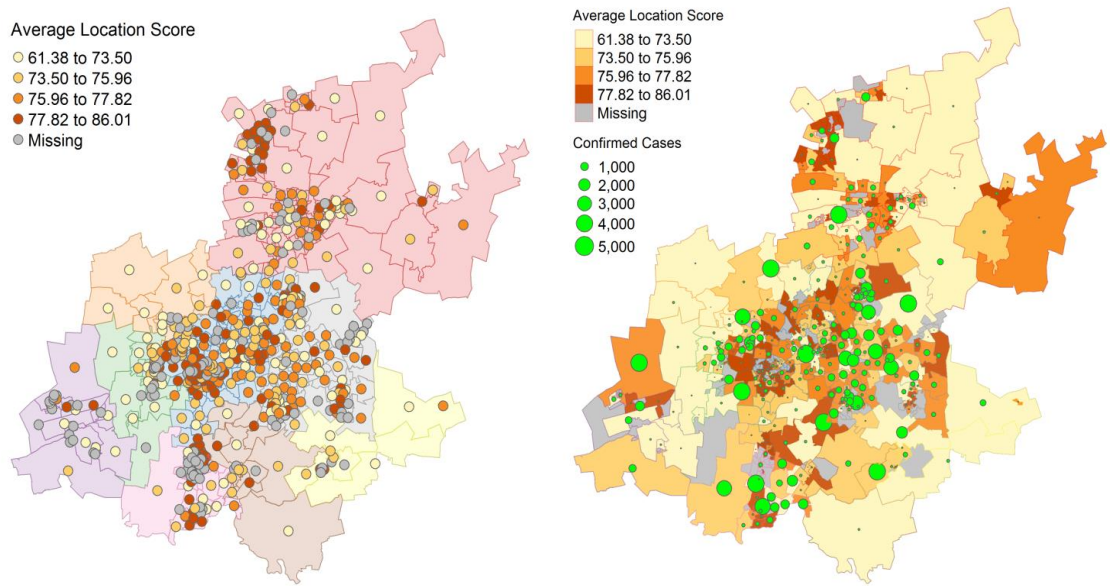

Figure 5: The spatial distribution of the average location score of the confirmed cases in each municipality of Gauteng.

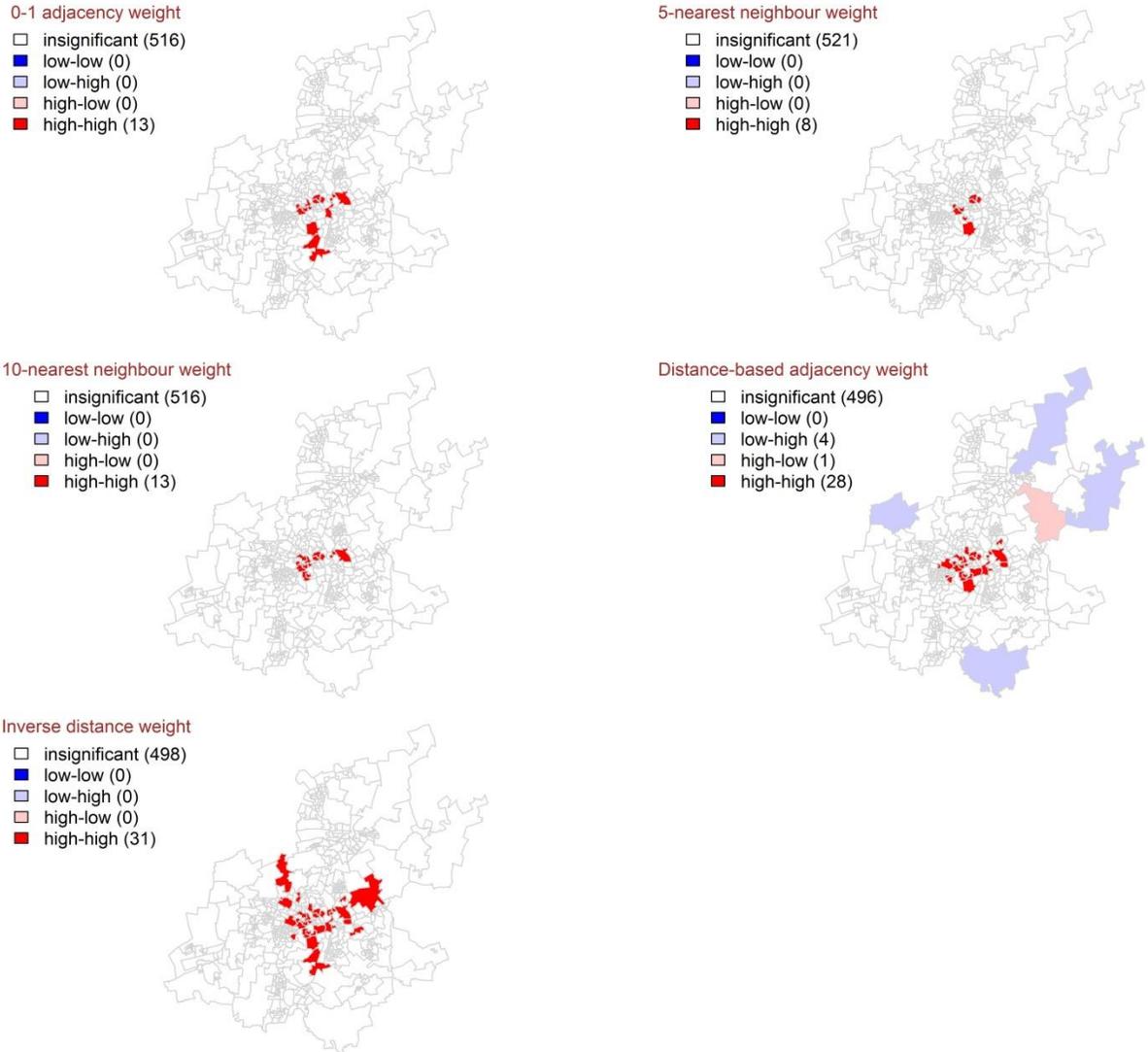

*Figure 6: The LISA maps created based on various weight matrices for COVID-19 confirmed cases of Gauteng at ward level. The values in the parenthesis are the number of wards fallen within that specified cluster/outliers.*

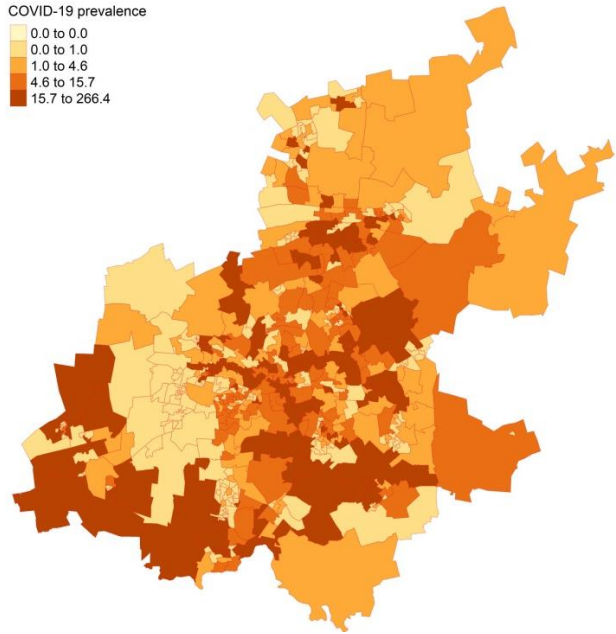

*Figure 7: The spatial distribution of COVID-19 prevalence in Gauteng province.*

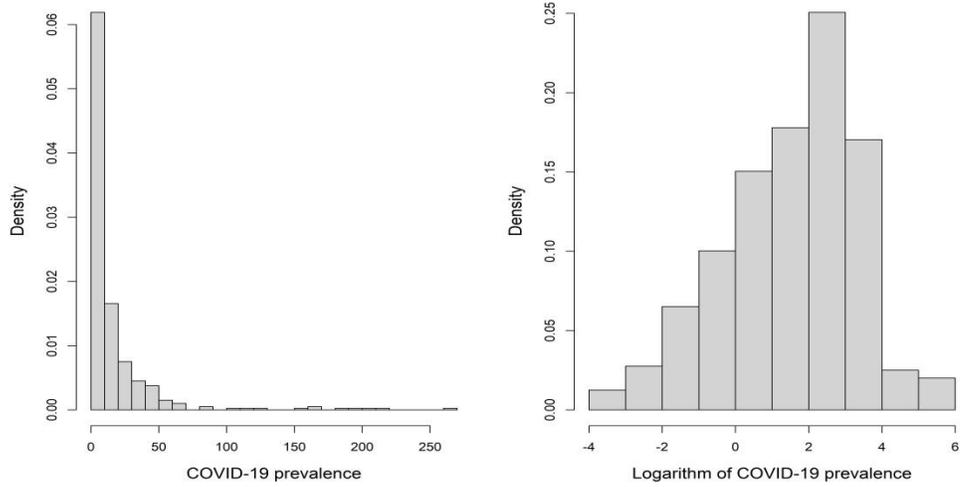

*Figure 8: The histograms of COVID-19 prevalence and its logarithm.*

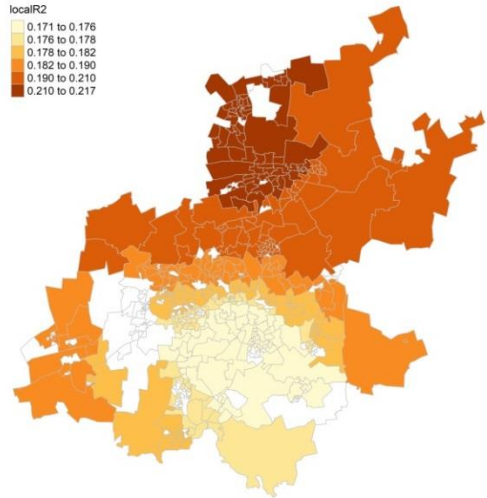

*Figure 9: The spatial distribution of the $R^2$ of the GWR model fitted on the prevalence of the COVID-19 of Gauteng's wards.*

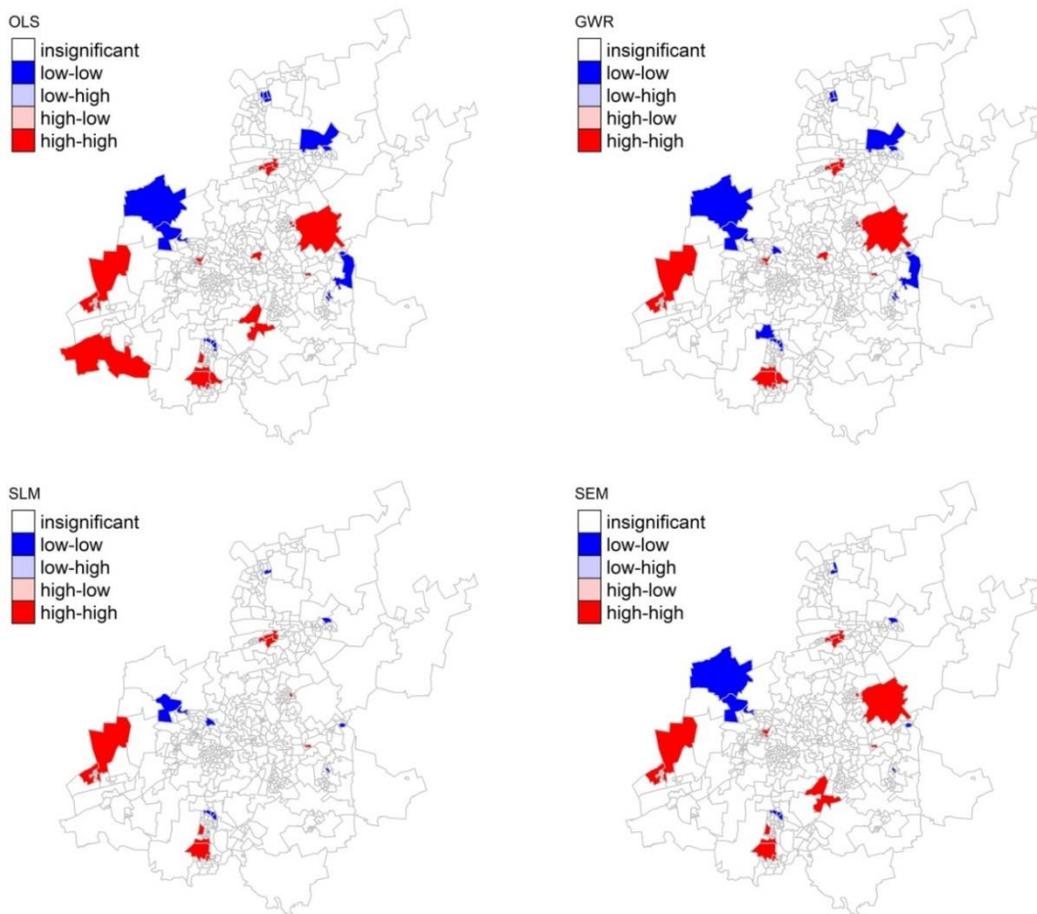

*Figure 10: The LISA maps created for the residuals of the various regression models fitted on the prevalence of the COVID-19 of Gauteng's wards.*

21